\documentclass[aps,prd,english,superscriptaddress,11pt,notitlepage]{revtex4}
	\usepackage[colorlinks=true, a4paper=true, pdfstartview=FitV,
linkcolor=blue, citecolor=blue, urlcolor=blue]{hyperref}

\usepackage{amsmath}
\usepackage{amssymb}
\usepackage{graphicx}
\usepackage{hhline}
\usepackage{textcomp}
\makeatletter
\usepackage{babel}
\newcommand{\bea}{\begin{eqnarray}}
\newcommand{\eea}{\end{eqnarray}}

\newcommand{\be}{\begin{equation}}
\newcommand{\ee}{\end{equation}}
\usepackage[active]{srcltx}
\begin{document}


 \title{The Dielectric Skyrme model}
 
 \author{C. Adam}
\affiliation{Departamento de F\'isica de Part\'iculas, Universidad de Santiago de Compostela and Instituto Galego de F\'isica de Altas Enerxias (IGFAE) E-15782 Santiago de Compostela, Spain}
\author{K. Oles}
\affiliation{Institute of Physics,  Jagiellonian University, Lojasiewicza 11, Krak\'{o}w, Poland}
\author{A. Wereszczynski}
\affiliation{Institute of Physics,  Jagiellonian University, Lojasiewicza 11, Krak\'{o}w, Poland}

\begin{abstract}
We consider a version of the Skyrme model where both the kinetic term and the Skyrme term are multiplied by field-dependent coupling functions. For suitable choices, this "dielectric Skyrme model" has static solutions saturating the pertinent topological bound in the sector of baryon number (or topological charge) $B=\pm 1$ but not for higher $|B|$.  This implies that higher charge field configurations are unbound, and loosely bound higher skyrmions can be achieved by small deformations of this dielectric Skyrme model. We provide a simple and explicit example for this possibility. 

Further, we show that the $|B|=1$ BPS sector continues to exist for certain generalizations of the model like, for instance, after its coupling to a specific version of the BPS Skyrme model, i.e., the addition of the sextic term and a particular potential. 
\end{abstract}

\maketitle

\section{Introduction}
The Skyrme model \cite{skyrme1}-\cite{skyrme3} is one particular proposal for a low-energy effective field theory (EFT) of strong-interaction physics \cite{thooft}-\cite{witten2} and, in particular, for the description of baryons and nuclei. Its primary fields are mesons, whereas baryons and nuclei emerge as topological solitons (Skyrmions) \cite{manton-book}-\cite{rho-ma-book} supported by the model. The Skyrme model incorporates many nontrivial features of low-energy QCD (baryon number conservation, chiral symmetry and its breaking, current algebra results, \ldots) in a completely natural way. It also reproduces some quantitative properties of nucleons and several light nuclei with reasonable success \cite{anw}-\cite{hal2}. Several shortcomings, however, impede its use as a general and quantitatively precise EFT of nuclear physics. Two major problems of the model are the too large binding energies of higher-charge Skyrmions and the absence of alpha-particle clusters inside them. Physical nuclei have rather small binding energies (always below 1\%) and frequently possess an alpha-particle substructure. 
The Skyrme model permits many generalizations, e.g., the addition of more terms \cite{castill} and the inclusion of further meson fields \cite{nappi1} -\cite{bando}, and some of these generalizations allow to significantly alleviate these shortcomings \cite{BPS-Sk}-\cite{vec-Sk-3}. 

The model proposed in the present letter is mainly motivated by the first problem (the too large binding energies), because the theory of topological solitons provides simple and systematic methods to search for models with small binding energies.  A systematic method to predict alpha-particle substructures for the Skyrmions of a particular model, on the other hand, is currently not known. To answer this question, at present full numerical calculations of higher charge solitons are required \cite{vec-Sk-3}. A better qualitative understanding of the formation of substructures within Skyrmions would certainly be desirable.   

The original version of the Skyrme model restricts the field content to pions and is given by the following lagrangian density
\be \label{orig-sk}
\mathcal{L}=\mathcal{L}_2+\mathcal{L}_4 +\mathcal{L}_0
\ee
where
\be
\mathcal{L}_2= c_2 \mbox{Tr} \; \partial_\mu U \partial^\mu U^\dagger, \;\; \mathcal{L}_4=c_4 \mbox{Tr} ([R_\mu, R_\nu])^2 
\ee
are the kinetic (Sigma model or Dirichlet) term and the Skyrme term, respectively. Here, the field $U$ takes values in SU(2), and $R_\mu = \partial_\mu U U^{-1}$ is the right-invariant current. Further, $\mathcal{L}_0 = -c_0\, \mathcal{U}(\mbox{Tr}\, U)$ is a potential term, where a frequent choice is the pion mass potential $\mathcal{U}_\pi = (1/2) \mbox{Tr}\,(\mathbb{I}- U)$. Finally, the $c_i$ are dimensionful coupling constants.

In the small-field limit, which is relevant for large distances, the Dirichlet term quadratic in the pion field dominates and induces attractive channels between Skyrmions. That is to say, there exist certain relative orientations between individual $B=1$ Skyrmions (modeling nucleons) such that they attract each other, and they may be arranged in arrays such that all nearest neighbors attract. The formation of bound states, i.e., the existence of higher charge Skyrmions with rest energies below the total mass of their constituents is, therefore, expected, and this is indeed what happens. If the pion mass term quadratic in the pion field is included, then the attractive forces change from power-law to exponential, but bound states still form. The resulting binding energies are, in fact, much larger than the binding energies of physical nuclei (see, e.g., \cite{manton-book}).   

The Skyrme model (\ref{orig-sk}) permits many generalizations. First of all, more general choices for the potential apart from the pion mass term are possible. Secondly, terms with higher powers of derivatives may be added. Among these, a particular term of a sixth power in first derivatives is singled out,
\be\label{sextic}
\mathcal{L}_6= -(24\pi^2)^2 c_6 \mathcal{B}_\mu \mathcal{B}^\mu ,
\ee
because this term is still quadratic in time derivatives and leads to a standard hamiltonian. Here, $\mathcal{B}^\mu$ is the baryon current,
\be \label{top-curr}
\mathcal{B}^\mu = \frac{1}{24\pi^2} \epsilon^{\mu \nu \rho \sigma} \mbox{Tr} \; R_\nu R_\rho R_\sigma, \;\;\; B= \int d^3 x \mathcal{B}^0 ,
\ee
which allows to calculate the baryon number (topological charge) $B$. Thirdly, in addition to the pions, further fields may be included in the model. 

Given this vast landscape of possible generalizations, arbitrarily choosing a model inside it and calculating its higher charge Skyrmions does not seem to be an efficient strategy for finding models with low binding energies. 
Topological soliton models like the generalized Skyrme models, however, allow to find nontrivial topological energy bounds \cite{bogom}, \cite{Fad}. For Skyrme models in Minkowski space-time, they are always exactly linear in the topological charge (baryon number) \cite{DH}, \cite{gen-Sk-bounds}.  Further, so-called Bogomolny equations can be found \cite{bogom}, which imply that the corresponding bound is saturated. These bounds and their Bogomolny equations are valuable tools in the search for Skyrme models with small binding energies. In a first step, certain submodels (so-called BPS models) must be identified which possess both a topological bound and nontrivial solutions (BPS Skyrmions) saturating the bound. The existence of the BPS Skyrmions implies that the binding energies in the model are either zero or negative. Models with small binding energies can then be constructed by certain "small" deformations of the BPS submodels. 

More concretely, a BPS submodel may either support BPS Skyrmions with arbitrary baryon number. Their energies are then exactly linear in $B$ and the resulting binding energies are zero. This is the case of the BPS Skyrme model \cite{BPS-Sk}, consisting of the sextic term (\ref{sextic}) and an arbitrary potential, or of the Skyrme model with an infinite tower of vector mesons \cite{vec-Sk-1}, obtained from a dimensional reduction of a higher-dimensional Yang-Mills theory.
The other possibility is that the BPS submodel has a BPS Skyrmion only in the $B=\pm 1$ sector. Higher $B$ solutions are then unstable, and the resulting "binding energies" are negative. In other words, the induced forces between the $B=1$ BPS Skyrmions are always repulsive. This is the case of the BPS submodel discovered by D. Harland \cite{DH}, and the model investigated in the present letter also belongs to this class. We remark that for BPS submodels based only on pion fields, either the absence of the Dirichlet term or its suppression in the small-field limit (e.g., by making it effectively higher than second order, or by enhancing other terms) is a necessary condition, as follows from the attractive forces induced by this term. Any deformation to a realistic near-BPS model must correct this behavior in the region of small pion fields.

\section{The dielectric Skyrme model}
We consider the static energy functional of the minimal Skyrme model (\ref{orig-sk}),
\be
E=E_2^d+E_4^d,  \label{model}
\ee
consisting of the kinetic term (Dirichlet energy),
\be
E_2^d=\int_{\mathbb{R}^3} -\frac{f^2}{2} \mbox{ Tr } (R_i R_i) d^3 x,
\ee
and the quartic Skyrme term
\be
E_4^d=\int_{\mathbb{R}^3} -\frac{1}{16 e^2} \mbox{ Tr } ([R_i, R_j] [R_i,R_j]) d^3 x.
\ee
In contrast to the standard case, however, we assume that  instead of the coupling constants $c_2$ and $c_4$ now we have field-dependent coupling functions $f$ and $e$. This is explicitly indicated by the index $d$ (dielectric). In particular, we assume that  they are functions of the trace of the Skyrme field $\mbox{Tr } U$,
implying that the isospin symmetry remains unbroken. Equivalently, using the standard parametrization (here $\vec\tau$ are the Pauli matrices)
\be
U=\exp (i \xi (\vec{x}) \vec{\tau} \cdot \vec{n}(\vec{x})), \label{U}
\ee
of the Skyrme field in terms of a profile function $\xi$ and an isospin unit vector $\vec n$, the coupling functions only depend on the profile,
$
f=f(\xi), \; e=e(\xi).
$

\subsection{Topological bound and Bogomolny equations}

To find the pertinent topological bound we follow the standard method based on the three eigenvalues $\lambda^2_i$ of the strain tensor $D_{ij}=-\frac{1}{2} \mbox{ Tr } (R_i R_j)$ \cite{NM-lambda}. Then, the static energy can be rewritten as
\bea
E^d&=&\int_{\mathbb{R}^3} \left[ f^2\left( \lambda^2_1+\lambda^2_2+\lambda^2_3\right) +  \frac{1}{e^2}\left(\lambda^2_1\lambda^2_2+\lambda^2_2\lambda^2_3 + \lambda^2_3\lambda^2_1\right) \right]d^3 x \nonumber \\
&=& \int_{\mathbb{R}^3} \left(  (f \lambda_1 \pm \frac{1}{e}   \lambda_2  \lambda_3)^2 + ( f\lambda_2 \pm \frac{1}{e} \lambda_3  \lambda_1)^2 + ( f\lambda_3 \pm  \frac{1}{e} \lambda_1 \lambda_2)^2 \right) d^3 x \mp 6 \int_{\mathbb{R}^3} \frac{f}{e} \lambda_1 \lambda_2 \lambda_3 d^3 x  \nonumber \\
& \geq & 6 \left| \int_{\mathbb{R}^3} \frac{f}{e}  \lambda_1 \lambda_2 \lambda_3 d^3 x \right| = 12 \pi^2 \left\langle\frac{f}{e} \right\rangle |B|.
\eea
where $\left\langle \mathcal{F} \right\rangle$ is the average value of a target space function $\mathcal{F}$ over the whole $\mathbb{S}^3$ target space. The last step follows from the fact that the baryon density 
$\mathcal{B}_0$ is just 
\be
\mathcal{B}_0 = \frac{1}{2\pi^2} \lambda_1 \lambda_2 \lambda_3.
\ee

The bound is saturated if and only if the following {\it dielectric} Bogomolny equations hold,
\be
 f \lambda_1 \pm \frac{1}{e} \lambda_2  \lambda_3 =0, \;\;  f \lambda_2  \pm  \frac{1}{e} \lambda_1\lambda_3 =0, \;\;  f\lambda_3 \pm   \frac{1}{e}  \lambda_1  \lambda_2 =0, \label{BOG}
\ee
which after a straightforward manipulation result in
\be
\lambda_1^2=\lambda_2^2=\lambda_3^2= e^2 \left( \xi \right) f^2 \left( \xi \right). \label{lambda}
\ee
In the standard case, where $f$ and $e$ are constant, the Bogomolny equations can not be satisfied for non-zero $B$ in $\mathbb{R}^3$ space. Indeed, then all eigenvalues must be constant which contradicts the finiteness of the energy for solutions of the Bogomolny equations. 
Here, on the other hand, the r.h.s. of (\ref{lambda}) is a function of  $\xi$, which can tend to 0 for $|\vec{x}| \to \infty$.
If we require that the Skyrme field $U$ approaches the perturbative vacuum $U=\mathbb{I}$ (i.e., $\xi \to 0$) for $|\vec{x}| \to \infty$, then 
 at least one of the two coupling functions, $e$ or $f$, must approach zero at $\xi =0$. We shall find that the solution presented in the next section automatically obeys this requirement.

\subsection{$B=1$ BPS solution}

It is known that for $\mathbb{R}^3$ base space the Bogomonly equations (\ref{lambda}) admit a topologically nontrivial solution if 
\be
ef = \frac{1}{2r_0} \mbox{ Tr } (\mathbb{I}-U) \label{cond}
\ee
where $r_0 >0$ (we chose the plus sign). In fact, this result was first obtained in the context of a BPS submodel considered by D. Harland \cite{DH}, consisting of the Skyrme term and the particular potential $\mathcal{U}_4=(\mbox{Tr } (\mathbb{I}-U))^4$ . The corresponding Bogomolny equations are {\it identical} to (\ref{lambda}) after the identification $(ef)^4 = \mathcal{U}_4$ and have solitonic solutions in the $B=\pm 1$ topological sectors. Skyrmions with higher values of the baryon charge do not obey the Bogomolny equations and, therefore, do not saturate the pertinent topological bound. 

Following \cite{DH}, the charge $B=1$ solution of our model can be easily found. We assume a natural spherical symmetry provided by the hedgehog ansatz $\xi = \xi (r)$, $\vec n = (\sin \theta \cos \phi , \sin\theta \sin \phi ,\cos \theta)$ in spherical polar coordinates $(r ,\theta ,\phi)$. The unit three component isovector $\vec{n}$ can be expressed via the stereographic projection by a complex field $u$
\be
\vec n = \frac{1}{1+|u|^2} \left(2 \Re(u), 2\Im(u),1-|u|^2\right). \label{n}
\ee
where $u(\theta, \phi)=\tan \frac{\theta}{2}e^{i\phi}$. Then the eigenvalues are
\be
\lambda_1^2=\xi_r^2, \;\;\; \lambda_2^2=\lambda_3^2=\frac{\sin^2 \xi }{r^2} .
\ee
The equality of the $\lambda_i^2$ leads to a first order ODE 
\be
\xi_r=-\frac{\sin \xi}{r}.
\ee
This should be completed with the third equality in (\ref{lambda}) 
\be
\xi_r=-\frac{1}{r_0} (1-\cos \xi) .
\ee
These two equations have a common solution
\be
\xi=2\arctan \frac{r_0}{r} \label{profile}
\ee
which interpolates between $\pi$ and 0 as $r$ changes from $0$ to infinity. This finally constitutes a BPS Skyrmion with unit topological charge. We remark that solution (\ref{profile}) also coincides with the solution found in a
background field
deformation of the Skyrme model \cite{LF}.

To compute the energy of the $B=1$ solution we must decide how the coupling functions depend on the target space coordinate $\xi$. This is, however, arbitrary provided the condition (\ref{cond}) is fulfilled. Here we consider the following possibility
\be
e=\frac{1}{r_0^\alpha} (1-\cos \xi)^\alpha, \;\;\; f=\frac{1}{r_0^{1-\alpha}} (1-\cos \xi)^{1-\alpha}
\ee
with the parameter $\alpha \in [0,1]$. Then using the explicit formula for the target space average integral
\be
 \left\langle \frac{f}{e} \right\rangle = \frac{2}{\pi} \int_0^\pi   \frac{f}{e}  \sin^2 \xi d\xi =  \frac{2}{\pi} \frac{1}{r_0^{1-2\alpha}} \int_0^\pi  \sin^2 \xi (1-\cos \xi)^{1-2\alpha} d\xi =  \frac{2\cdot 4^{1-\alpha}}{\sqrt{\pi}} \frac{1}{r_0^{1-2\alpha}} \frac{\Gamma \left[ \frac{5}{2}-2\alpha\right]}{\Gamma[4-2\alpha]}
\ee 
finally, we get 
\be
E^d=12\pi^2  \frac{2\cdot 4^{1-\alpha}}{\sqrt{\pi}} \frac{1}{r_0^{1-2\alpha}} \frac{\Gamma \left[ \frac{5}{2}-2\alpha\right]}{\Gamma[4-2\alpha]} .
\ee

As two extremal cases we take $\alpha=0$ and $\alpha=1$ (the case $\alpha =0$ was considered already in \cite{na-ol}, in a slightly different context). In the first possibility, the Skyrme term is multiplied by a constant while the first dielectric function $f=(1-\cos \xi)/r_0$. Then, the energy of the unit BPS Skyrmion is $E= 12\pi^2/r_0$. On the other hand, if $\alpha=1$ then only the Skyrme part is multiplied by a dielectric function, $e=(1-\cos \xi)/r_0$. Now, the energy is $E=24\pi^2 r_0$.

The restrictive form of the Bogomolny equations excludes solutions with higher values of the topological charge for the same model (the same dielectric functions $e$ and $f$). The arguments presented in \cite{DH} likewise hold for the dielectric Skyrme model. The result is that higher charge Skyrmions are not BPS solitons and have energies higher than $B\cdot E(B=1)$. Thus, they are energetically unstable towards a decay into a collection of separated charge one BPS Skyrmions. To conclude, there are no stable $B>1$ Skyrmions. 

\subsection{A dielectric near-BPS Skyrme model}
The behavior (\ref{cond}) of the dielectric functions is not phenomenologically acceptable close to the vacuum $\xi =0$. A rather obvious proposal for a realistic near-BPS model is a deformation which changes (\ref{cond}) to a nonzero value for $\xi \to 0$ but leaves it essentially untouched for sufficiently large $\xi$. This proposal has the twofold advantage that, $i)$, it recovers the correct small-field limit and, $ii)$, it should provide small binding energies, because small-field regions  only make small contributions to the total energy.  For a fully realistic model one probably prefers smooth dielectric functions $f(\xi)$ and $e(\xi)$, but for the less ambitious goal of finding estimates for the binding energies of near-BPS models, continuous functions are sufficient. Concretely, we shall assume $e=1$, whereas $f$ is given by the expression resulting from (\ref{cond}) for $\xi \in [\xi_*,\pi]$, but by the constant value $f_* = f(\xi_*)$ for the near-vacuum region $\xi \in [0,\xi_*]$.  For our specific example we choose $\xi_* = (\pi/6)$ which is small but not very small. Further, the size parameter $r_0$ is irrelevant for our energy considerations, therefore we choose $r_0=1$.
That is to say, we choose de dielectric Skyrme model with the dielectric functions
\be \label{near-BPS-model}
e = 1 \, , \quad f = \left\{ \begin{array}{ccc} 1-\cos\xi & \ldots &  \; \xi \in [\xi_*,\pi]\\ f_* \equiv 1- \cos \xi_* = 1 - \frac{\sqrt{3}}{2} &\ldots & \; \xi \in [0,\xi_*] \end{array} \right. \, , \quad \xi_* \equiv \frac{\pi}{6} .
\ee 
For $\xi \in [\xi_*,\pi]$, the contribution $E^>$ to the energy is given by the accordingly restricted BPS bound,
\bea
E^> =E^>_{\rm BPS} &=& 12 \pi^2 \langle 1-\cos \xi\rangle_> = 24 \pi \int_\frac{\pi}{6}^\pi d\xi \sin^2\xi (1-\cos\xi) \nonumber \\ &=& \pi (10\pi + 1 + 3\sqrt{3} ) \simeq 118.162 \, .
\eea
For $\xi \in [0,\xi_*]$, the contribution to the BPS bound is simply given by the accordingly restricted Skyrme-Faddeev bound,
\be
E^<_{\rm BPS} = 24\pi f_* \int_0^\frac{\pi}{6} d\xi \sin^2 \xi = \pi f_* (2\pi - 3\sqrt{3}) \simeq 0.4576 \, .
\ee
For the contribution $E^<$ to the true energy, we should, in principle, find the hedgehog solution of the minimal Skyrme model (with coupling constants $e=1$ and $f=f_*$) in the region $\xi \in [0,\xi_*]$ with the corresponding boundary conditions. But if we only want to find an upper bound for the binding energy, then an upper bound $E_{\rm b}^< >E^<$ is sufficient. Such upper bounds can be found by inserting certain trial functions instead of the true hedgehog solution into the energy functional. One first possibility is to use the same BPS solution (\ref{profile}), but it turns out that this is a lousy approximation, because it has the wrong large $r$ behavior. The solution of the minimal Skyrme model behaves like $r^{-2}$ for large $r$, so the simplest possible trial function with this behavior is
\be
\xi^< (r) = \xi_* \frac{r_*^2}{r^2}  \quad \ldots \quad r > r_* , \quad r_* = \frac{1}{\tan\frac{\pi}{12}} = \frac{1}{2-\sqrt{3}}.
\ee
Here, $r_*$ is the radius where the BPS solution (\ref{profile}) takes the value $\xi_*$, such that the BPS solution for $r\le r_*$ and the trial function $\xi^<$ for $r>r_*$ together define a continuous function. Its first derivative is no longer continuous (has a finite jump at $r_*$), but this is sufficiently regular for our energy estimates. The numerical values of our parameters are
\be
f_* \simeq 0.1340 \, , \quad r_* \simeq 3.7320 \, , \quad C\equiv \xi_* r_*^2 \simeq 7.2928 .
\ee
For the upper energy bound we finally get 
\be
E_{\rm b}^< = E_{{\rm b},2}^{<} + E_{{\rm b},4}^{<} \equiv 4\pi \int_{r_*}^\infty dr r^2 (\mathcal{E}_{{\rm b},2} + \mathcal{E}_{{\rm b},4} )
\ee
where 
\bea
\mathcal{E}_{{\rm b},2} &=& f_*^2 \left( 4\frac{C^2}{r^6} + 2 \frac{\sin^2 (C/r^2)}{r^2} \right) \\
\mathcal{E}_{{\rm b},4} &=& 8\frac{C^2}{r^8} \sin^2 \frac{C}{r^2} + \frac{\sin^4 (C/r^2)}{r^4}.
\eea
Performing the integrations numerically, this leads to
\bea
E_{{\rm b},2}^{<} &=& 4\pi f_*^2 \cdot 2.0203 = 0.4559 , \\
E_{{\rm b},4}^{<} &=& 4\pi \cdot 0.2359 = 0.2359 , \\
E_{\rm b}^< &=& E_{{\rm b},2}^{<} + E_{{\rm b},4}^{<} = 0.6918 . 
\eea
For the relative binding energies we, therefore, get the following {\em upper bound},
\be
\frac{\Delta E(B)}{E(B) } \equiv \frac{BE(1) -E(B)}{E(B) }\le \frac{E_{\rm b}^< - E^<_{\rm BPS}}{E^>_{\rm BPS}+E^<_{\rm BPS}} \simeq \frac{0.235}{118.6} = 0.00198 ,
\ee
where $E(B)$ is the energy of a skyrmion with baryon number $B$.
In other words, relative binding energies for the dielectric Skyrme model defined by the dielectric functions of Eq. (\ref{near-BPS-model}) must always be below 0.2\%. This example demonstrates that the dielectric Skyrme model not only allows to find submodels with small binding energies. It provides, in fact, an extremely simple and natural mechanism to construct such models with arbitrarily small binding energies.

\section{Inclusion of the sextic term and potential }
\subsection{Dielectric BPS Skyrme model}

If generalizations of the Skyrme model at most quadratic in time derivatives are considered, which still lead to a standard hamiltonian, then two possible terms may be added. Namely, the sextic term with static energy
$E_6= g^2 \pi^2 \mathcal{B}_0^2$, and a non-derivative term, i.e., a potential $E_0\equiv \mathcal{U}$. The two terms together form the so-called BPS Skyrme model 
\be
E_{BPS}=E_6+E_0
\ee
Here $g$ is a positive constant.

This model is interesting because some of its properties coincide with several relevant features of nuclear matter (atomic nuclei). First of all, it is a BPS theory where the corresponding Bogomolny equation has solitonic solutions in {\it any} topological sector. Thus, stable BPS solitons with arbitrary values of the baryon charge exist. As a consequence, the model provides zero binding energies at the classical level. Rather realistic physical binding energies can be obtained already within the model, if some natural classical and quantum corrections are included \cite{PRL}. Secondly, the static energy functional enjoys a large symmetry group, i.e., the volume preserving diffeomorphisms. This means that the energy of a BPS soliton does not depend on its shape and is constant provided the volume remains unchanged. In fact, this symmetry is the symmetry of a liquid if the surface energy is negligible. Furthermore, the BPS Skyrme Lagrangian describes a perfect fluid. This reproduces at the field theoretic level the liquid drop model.

Let  us now promote the constant $g$ to a function of the target space variable, $\mbox{Tr } U$. We do not consider any dielectric function for $E_0$, as any functional dependence may incorporated into the potential. The resulting dielectric version of the BPS Skyrme model still possesses a non-empty self-dual sector. To see this, we compute the corresponding topological bound
\bea
E_{BPS}^d&=&\int_{\mathbb{R}^3} \left( \frac{g^2}{4\pi^2} \lambda^2_1\lambda^2_2\lambda^2_3 + \mathcal{U} \right)d^3 x = \int_{\mathbb{R}^3}  \left( \frac{g}{2\pi} \lambda_1\lambda_2\lambda_3 \pm \sqrt{\mathcal{U}}   \right)d^3 x   \mp \frac{1}{\pi} \int_{\mathbb{R}^3} g\sqrt{\mathcal{U}} \lambda_1 \lambda_2 \lambda_3 d^3 x  \nonumber \\
& \geq & \frac{1}{\pi} \left| \int_{\mathbb{R}^3} g\sqrt{\mathcal{U}}   \lambda_1 \lambda_2 \lambda_3 d^3 x \right| = 2 \pi \left\langle g\sqrt{\mathcal{U}}  \right\rangle |B|.
\eea
The bound is saturated if and only if the following Bogomolny equation is obeyed
\be
\frac{g}{2\pi} \lambda_1\lambda_2\lambda_3 \pm \sqrt{\mathcal{U}} =0. \label{BOG-BPS}
\ee
This can be transformed into the Bogomolny equation of the standard (non-dielectric) BPS Skyrme model by introducing a new potential $\tilde{\mathcal{U}}=g^2\mathcal{U}$. In other words, the dielectric function can always be incorporated into the potential. As a consequence, the {\it SDiff} invariance of the BPS solutions survives. To conclude, the dielectric generalization of the BPS Skyrme model does not change the main qualitative properties of the model. In realistic applications, the functions $g$ and $\mathcal{U}$  should, of course, be constrained by experimental data. 

\subsection{$B=1$ BPS solution of the generalized dielectric Skyrme model}
Owing to the large freedom in the (dielectric) BPS Skyrme model, it is possible to find such functions $g$ and $\mathcal{U}$ that the Bogonolny equation {\it shares} some solutions with the Bogomolny equations for the dielectric minimal Skyrme model (\ref{BOG}).

In the unit charge sector, a Skyrmion again can be obtained by the hedgehog ansatz. Thus, eq. (\ref{BOG-BPS}) leads to 
\be
\frac{1}{2\pi} \frac{g(\xi) \sin^2 \xi  \xi_r}{ r^2} = - \sqrt{\mathcal{U}}
\ee 
Assuming that the solution should be of the form (\ref{profile}) we can find that 
\be
\frac{\sqrt{\mathcal{U}}}{g}=\frac{1}{2\pi r_0^3} (1-\cos \xi)^3 \equiv \frac{1}{16\pi r_0^3} \left( \mbox{Tr } (\mathbb{I}-U) \right)^3 \label{cond-2}
\ee
Thus, we have proven that the generalized dielectric Skyrme model 
\be
E^d=E_2^d+E_4^d+E_6^d+E_0
\ee
has a BPS unit charge solution provided the dielectric functions and the potential obey relations (\ref{cond}) and (\ref{cond-2}).  

Note that the addition of the minimal (dielectric) Skyrme model breaks the {\it SDiff} symmetry explicitly. Furthermore, higher charge Skyrmions are again unstable towards decay into separated BPS Skyrmions.
\section{Possible applications and conclusions }
A first, general observation is that there exists a rather large freedom in the construction of  near-BPS Skyrme models, where solitons form bound states with small binding energies. The dielectric Skyrme models proposed in the present letter constitute a new and interesting possibility for this phenomenon, which should be further explored. In particular, the inclusion of the sextic term (or the BPS Skyrme model) into the full BPS submodel is interesting, because this part of the complete model gives the leading behavior in the high density (pressure) regime \cite{nucl-eos}.  

Small classical binding energies require that the BPS property in the $B=1$ sector is weakly broken. This can be achieved by a deformation of the constraints on the dielectric functions and the potential (\ref{cond}), (\ref{cond-2}). 
For any deformation to a physically relevant near-BPS model we should impose that close to the perturbative vacuum ($\xi=0$) the dielectric functions tend to the non-zero vacuum values $f(\xi=0)=f_\pi$ and $e(\xi=0)=e_0$. 
But these conditions only affect the small-field regions which provide rather small contributions to the total energy of a soliton. In other words, if we choose deformations such that the coupling functions remain (almost) unchanged for large field values, then the resulting binding energies are expected to be small. We demonstrated in a concrete  and simple example that this is indeed the case, and the resulting binding energies can be made extremely small.
That is to say, the most natural deformations of the dielectric BPS Skyrme model, i.e., those which just recover the phenomenologically correct small-field limit, automatically provide very low binding energies, by construction.

A related question concerns possible physical interpretations or justifications for the (deformed) coupling functions. From an effective field theory point of view, they simply correspond to higher order terms in the field expansion which may be taken into account. They are not forbidden because they respect the relevant symmetries. Further, after the deformation, they also reproduce the correct small-field limit. Another possibility is to interpret the
 dielectric functions as in-medium coupling constants, in particular, $f$ as an in-medium pion decay constant. The same concerns the dielectric function $g$ and especially the potential $\mathcal{U}$, which in the small field-limit in vacuum should tend to the pion mass potential $m^2\mbox{Tr } (\mathbb{I}-U)$. 

The possible relation between (near)-BPS structures and in-medium properties of Skyrmions (in-medium Skyrme models, see, e.g.,  \cite{in-med1}-\cite{in-med4}) indicated above is an interesting observation which deserves a more profound investigation.

We remark that the condition (\ref{cond-2}) also allows to add the model $E_4+\mathcal{U}_4$ of \cite{DH}. 

\section*{Acknowledgements}
The authors acknowledge financial support from the Ministry of Education, Culture, and Sports, Spain (Grant No. FPA2017-83814-P), the Xunta de Galicia (Grant No. INCITE09.296.035PR and Conselleria de Educacion), the Spanish Consolider-Ingenio 2010 Programme CPAN (CSD2007-00042), Maria de Maetzu Unit of Excellence MDM-2016-0692, and FEDER. 

\end{document}